\documentclass{article}

\usepackage{arxiv}

\usepackage[utf8]{inputenc} 
\usepackage[T1]{fontenc}    
\usepackage{hyperref}       
\usepackage{url}            
\usepackage{booktabs}       
\usepackage{amsfonts}       
\usepackage{nicefrac}       
\usepackage{microtype}      
\usepackage{lipsum}
\usepackage{graphicx}
\usepackage[square,numbers]{natbib}

\title{CHAOTIC SADDLES IN A GENERALIZED LORENZ MODEL OF MAGNETOCONVECTION}

\author{
  Francis F.~Franco \\
   Department of Physics\\
   Technological Institute of Aeronautics (ITA)\\
   S\~ao Jos\'e dos Campos, SP, Brazil, 12228-900\\
   \texttt{francisferreirafranco@gmail.com} \\
   \And
 Erico L.~Rempel \\
  Departmanet of Mathematics\\
  Technological Institute of Aeronautics (ITA)\\
  S\~ao Jos\'e dos Campos, SP, Brazil, 12228-900\\
  \texttt{rempel@ita.br} \\
}

\begin{document}
\maketitle

\begin{abstract}
The nonlinear dynamics of a recently derived generalized Lorenz model (Macek and Strumik, Phys. Rev. E 82, 027301, 2010) of magnetoconvection is studied. A bifurcation diagram is constructed as a function of the Rayleigh number where attractors and nonattracting chaotic sets coexist inside a periodic window. The nonattracting chaotic sets, also called chaotic saddles, are responsible for fractal basin boundaries with a fractal dimension near the dimension of the phase space, which causes the presence of very long chaotic transients. It is shown that the chaotic saddles can be used to infer properties of chaotic attractors outside the periodic window, such as their maximum Lyapunov exponent.
\end{abstract}

\keywords{Chaos, Chaotic saddles, Lorenz model, Reduced model, Long chaotic transients}

\section{Introduction}
The study of simple models of thermal convection based on truncated solutions of hydrodynamic equations has been extremely popular since Lorenz's original work on deterministic nonperiodic flows \cite{lorenz1963}. The model was originally derived for the study of two-dimensional Rayleigh-B\'enard convection, i.e., thermal convection in a plane layer of fluid heated from below and cooled from above \cite{chandra}. Since then, sets of Lorenz-like equations have been obtained in different contexts, such as lasers \cite{haken1975}, dynamos \cite{knobloch1981,weiss1984}, magnetoconvection \cite{knobloch1992,macek2010}, chemical reactions \cite{poland1993} etc. In the present paper, we explore the generalized Lorenz model introduced by \citet{macek2010} in the context of magnetoconvection, where the interaction of an electrically conducting fluid with an imposed magnetic field is considered. The model follows Lorenz's original derivation of three nonlinear ordinary differential equations, adding a fourth equation for the magnetic field fluctuation and it was shown to exhibit hyperchaos \cite{macek2014,macek2018}. Although most previous analysis of Lorenz systems (and dynamical systems in general) focus on the asymptotic behaviour, after solutions have converged to an attractor, here we stress the importance of the initial transient dynamics and identify the role of nonattracting chaotic sets.
 
Nonattracting chaotic sets are subsets of the phase space of a dynamical system where a nonattracting chaotic trajectory can be found, with the corresponding properties of aperiodicity and sensitivity to initial conditions, as measured by a positive Lyapunov exponent \cite{hsu1988}. Being nonattracting means that neighbouring trajectories will wonder in the vicinity of a chaotic set for a finite time before they diverge from it, eventually converging to some coexisting attractor in the case of dissipative systems. Thus, the outcome is a transient or decaying chaotic behaviour. Usually, there is a special (fractal) set of initial conditions called {\em stable manifold} that converges to the nonattracting chaotic set in forward time, and a set that converges to it in reversed time dynamics, the {\em unstable manifold} \cite{hsu1988,sweet2000}. For that reason, nonattracting chaotic sets are also known as chaotic saddles, as they lie at the intersection of their stable and unstable manifolds, which are smooth surfaces in the phase space. Chaotic saddles are also known as chaotic repellors and are related to dynamical phenomena like chaotic scattering \cite{macau2002} and fractal basin boundaries \cite{battelino1988}.

Another important role played by chaotic saddles in dynamical systems is in global bifurcations known as crises \cite{grebogi1982}, where chaotic attractors undergo a sudden change in size or structural stability. In a boundary crisis, a chaotic attractor suddenly disappears, leaving a chaotic saddle in its place \cite{robert2000,chian2007}; in an interior crisis, a chaotic attractor suddenly enlarges after collision with a chaotic saddle \cite{szabo1994}; in a merging crisis, two or more chaotic attractors are united to form a larger attractor, where the former attractors are converted to chaotic saddles \cite{rempel2005}. In all types of crises, chaotic saddles affect the observable dynamics through the appearance of chaotic transients and/or crisis-induced intermittency \cite{romeiras1987,rempel2003,rempel2007}. Due to their close relation to crises, the properties of a chaotic saddle can be used to infer the properties of chaotic attractors generated at crises, such as their Lyapunov exponents and overall topology \cite{szabo1994,rempel2005,tel2015}. In this work, we exemplify this fact in the generalized Lorenz model by comparing the shape of the invariant sets and the value
of their maximum Lyapunov exponents for a chaotic saddle and a chaotic attractor in the phase space for different values of the control parameter. 
 
This paper is organized as follows. In section \ref{sec model}, the generalized Lorenz model of magnetoconvection is described; in section \ref{sec analysis}, the numerical nonlinear analysis of the model is presented; section \ref{sec conclusion} provides the conclusions.

\section{The Generalized Lorenz Model}
\label{sec model}
In this section we summarize the derivation of the generalized Lorenz model of magnetoconvection, which is provided in more details in \citet{macek2018}. Consider the two-dimensional motion of a conducting fluid between two horizontal plates separated by a height $h$ in the $z$ direction, with the temperature at the bottom plate kept higher than the temperature at the top plate and an imposed magnetic field in the horizontal direction ($x$). The evolution of the magnetized fluid is governed by the magnetohydrodynamic (MHD) equations

 \begin{eqnarray}
  \frac{D \textbf{v}}{Dt} &=&  - \frac{1}{\rho} \nabla \left(p + \frac{\textbf{B}^2}{2 \mu_{0}} \right) + \frac{\left( 
  \textbf{B} \cdot \nabla \right) \textbf{B}}{\mu_{0} \rho} + \nu \nabla^2 \textbf{v} + \rho \mathbf{g}, \\ 
 \frac{D \textbf{B}}{Dt} &=& \left( \textbf{B} \cdot \nabla \right) \textbf{v} + \eta \nabla^2 \textbf{B},\\
  \frac{D T}{D t}&=& \kappa \nabla^2 T,\\
     \nabla \cdot \textbf{B} &=& 0,
 \end{eqnarray}
where $\mathbf{v}$ denotes the velocity of the flow, $\rho$ is the mass density, $p$ the pressure, $\textbf{B}$ is the magnetic field, $\mu_{0}$ is the permeability of vacuum, $\nu$ is the kinematic viscosity, $\eta$ the magnetic resistivity, $\kappa$ the thermal conductivity of the fluid, $T$ is the temperature, $\mathbf{g}$ is the constant gravity and $D/Dt \equiv (\partial/ \partial t)+ \mathbf{v} \cdot \mathbf{\nabla}$. Additionally, consider the Boussinesq approximation, where the mass density is considered constant $\rho=\rho_0$, where $\rho_0$ is the density at the lower boundary, except in the buoyancy term ($\rho \mathbf{g}$), where $\rho$ is given by $\rho = \rho_{0}[1 -\beta (T-T_b)]$, where $\beta$ is the constant thermal expansion coefficient and $T_b$ is the temperature at the bottom plate.

By adopting a stream function for the flow velocity, a vector potential for the magnetic field and employing a one-mode Fourier representation, \citet{macek2014} obtained the following generalized Lorenz model 
\begin{eqnarray}
  \dot{x}&=&-\sigma x + \sigma y - \omega_{0} w,\label{model1}\\
  \dot{y}&=&-x z + r x -y,\\
  \dot{z}&=&x y - b z,\\
  \dot{w}&=&\omega_{0} x - \sigma_{m} w,\label{model4}
 \end{eqnarray}
where $x$ is the Fourier coefficient related to the stream function, $y$ and $z$ are the coefficients related to the temperature fluctuation and $w$ is the coefficient related to the magnetic vector potential. The overdot denotes derivative with respect to the normalized time $t'=(1+a^2) \kappa (\pi/h)^2t$, where $a$ is a parameter associated with the width $(h/a)$ of the convective rolls at the onset of convection. The other parameters are $b=4/(1+a^2)$, the Prandtl number $\sigma=\nu/\kappa$, the magnetic Prandtl number $\sigma_m=\eta/\kappa$, the normalized Rayleigh number $r=R_a/R_c$, where $R_a=g\beta h^3\delta T/(\nu\kappa)$ and  $R_c=(1+a^2)^3(\pi^2/a)^2$, and $\omega_0=\upsilon_{A_0}/\upsilon_0$ is responsible for the strength of the imposed magnetic field, with $\upsilon_0 = 16\pi^2\kappa/(abh\mu_0)$ and $\upsilon_{A_0}=B_0/(\mu_0\rho_0)^{1/2}$ is the Alfv\'en speed. When $\omega_0=0$ the traditional Lorenz model is obtained.

\section{Nonlinear Dynamics Analysis}
\label{sec analysis}

\begin{figure}[!htp]
\begin{center}
\includegraphics[width=7.0cm]{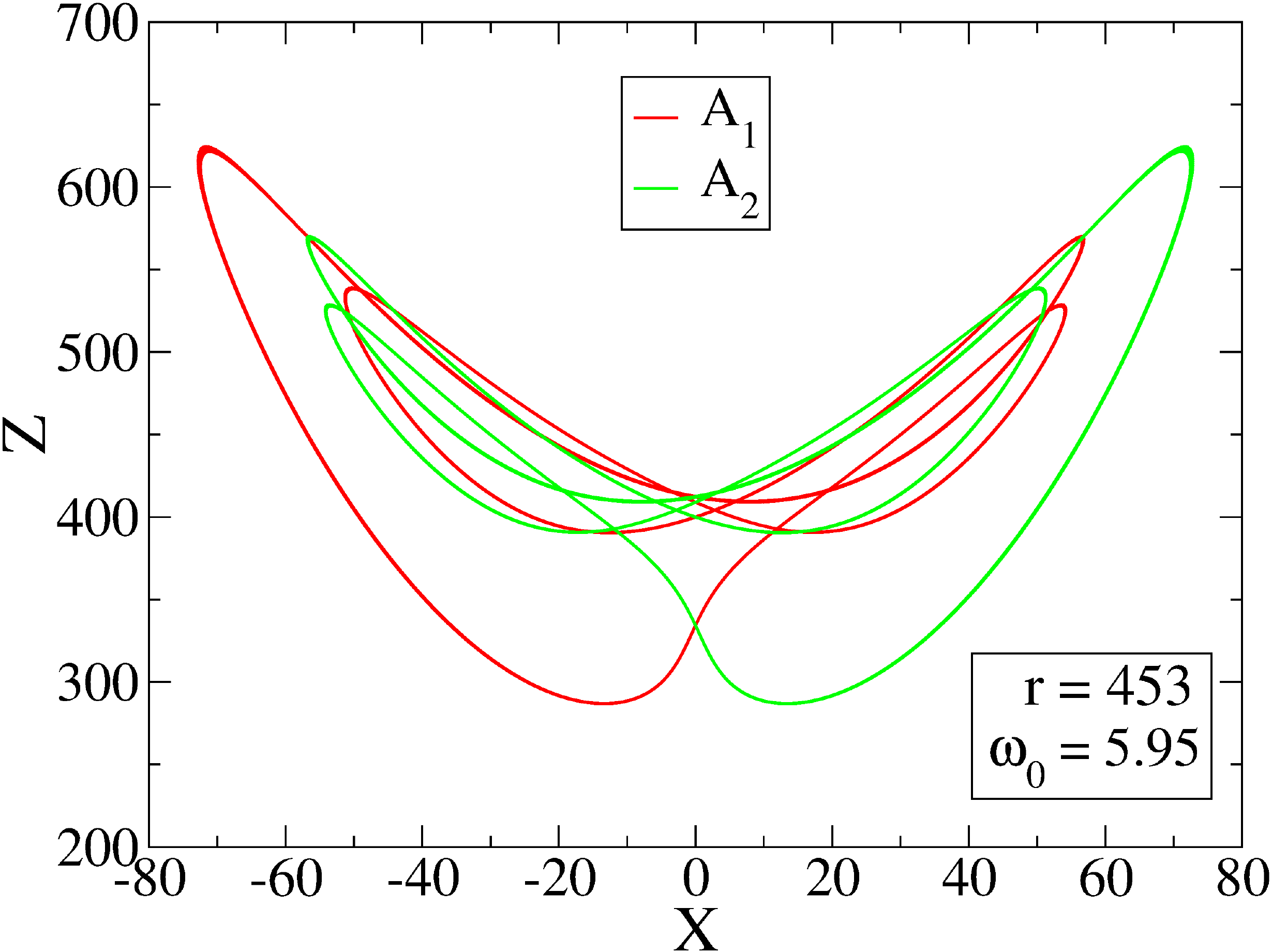}
\end{center}
\caption{Symmetry under reflection through the z axis shown in projection to the x,z plane for periodic attractors A1 (red) and A2 (green) for $\sigma=10$, $b=8/3$, $\sigma_{m}=0.1$, $\omega_{0}=5.95$ and $r=443$.}
\label{fig symmetry}
\end{figure}

Equations (\ref{model1})-(\ref{model4}) are solved with a fourth-order Runge-Kutta integrator. Just like the original Lorenz system, the generalized model exhibits symmetry under reflection through the $z$ axis, i.e., the system equations are reversible under $x \mapsto -x$, $y\mapsto-y$, $w\mapsto-w$. Figure \ref{fig symmetry} illustrates this symmetry, where a periodic attractor (red) and its symmetric counterpart (green) are plotted. Due to this property, the nonlinear evolution of an attractor under changes in control parameters is mimicked by its symmetric attractor, with identical bifurcations taking place simultaneously in different parts of the phase space. We refer to both attractors as $A_1$ and $A_2$.

\begin{figure}[!htp]
\begin{center}
\includegraphics[width=6.5cm]{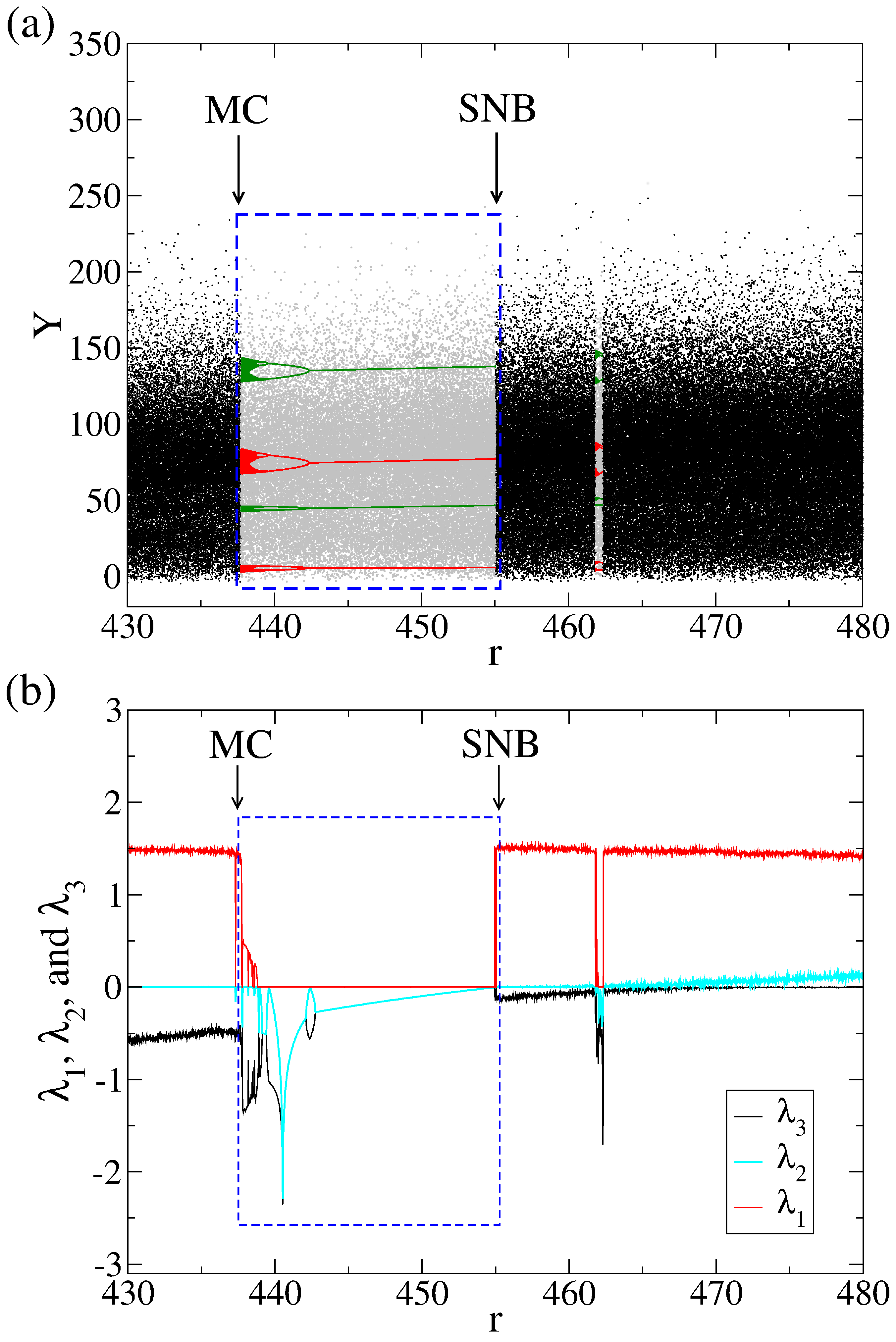}
\end{center}
\caption{(a) Bifurcation diagram showing the Poincar\'e points of state variable $y$ as a function of the control parameter $r$, with $\sigma=10$, $b=8/3$, $\sigma_m=0.1$ and $\omega_0 = 5.95$. The gray area represents the chaotic saddle $\Lambda_s$, red dots represent attractor $A_1$ and green dots represent attractor $A_2$; MC denotes merging crisis and SNB denotes saddle-node bifurcation. (b) The three largest Lyapunov exponents, $\lambda_1$, $\lambda_2$ and $\lambda_3$, for the attractors in (a) as a function of $r$.}
\label{fig bifdiag}
\end{figure}

We adopt the Poincar\'e map defined by $x=0$ with $\dot{x}>0$, thus a point is plotted everytime a solution crosses the hyperplane $x=0$ from ``left'' to ``right''. Figure \ref{fig bifdiag}(a) shows the bifurcation diagram of the generalized Lorenz model for $\sigma=10$, $b=8/3$, $\sigma_m=0.1$ and $\omega_0 = 5.95$, while $r$ is varied between 430 and 480. This set of parameter values was chosen following \citet{macek2014}. For each value of $r$, an initial condition is integrated until the orbit converges to an attractor, after which we start plotting the $z$ component of the Poincar\'e points in red dots. There is a period-2 periodic window starting with a saddle-node bifurcation (SNB) at $r \approx 455.47$. By reducing $r$, the period-2 attractor undergoes a flip bifurcation at $r\approx 442.35$, where its period in the Poincar\'e map duplicates, going from period-2 to period-4. As the reduced Rayleigh number $r$ is further decreased, a cascade of period-doubling (flip) bifurcations takes place, leading to a small chaotic attractor localized in two narrow bands. The two green lines inside the window represent the evolution of attractor $A_2$ in parallel with $A_1$. The grey dots represent the transient chaotic behaviour displayed by the trajectories before they converge to either $A_1$ or $A_2$ and were found with the sprinkler method \cite{hsu1988}. The chaotic transients are due to a chaotic saddle surrounding the attractors, and we denote this chaotic saddle by $\Lambda_s$. The window ends in a merging crisis (MC) at $r\approx  437.73$, where $A_1$ and $A_2$ simultaneously collide with the surrounding chaotic saddle and the three sets merge, leading to the formation of a large chaotic attractor. To the left of MC as well as to the right of SNB, the trajectories of $A_1$ and $A_2$ are united in a single large chaotic attractor, except in some narrow periodic windows, where the two attractors split again. Figure \ref{fig bifdiag}(b) displays the three largest Lyapunov exponents ($\lambda_1 > \lambda_2 > \lambda_3$) of the attracting sets in Fig. \ref{fig bifdiag}(a) and is a reproduction of Fig. 2 of \citet{macek2014}. Note that $\lambda_1$ suddenly drops to negative values at SNB, rises to positive values near MC and then, suddenly jumps to a much higher value at MC. There is an interval for $r>454.7$ where hyperchaos is found, with $\lambda_1>0$ and $\lambda_2>0$, a phenomenon first reported in this system by \citet{macek2014}.

\begin{figure}[!htp]
\begin{center}
\includegraphics[width=8.0cm]{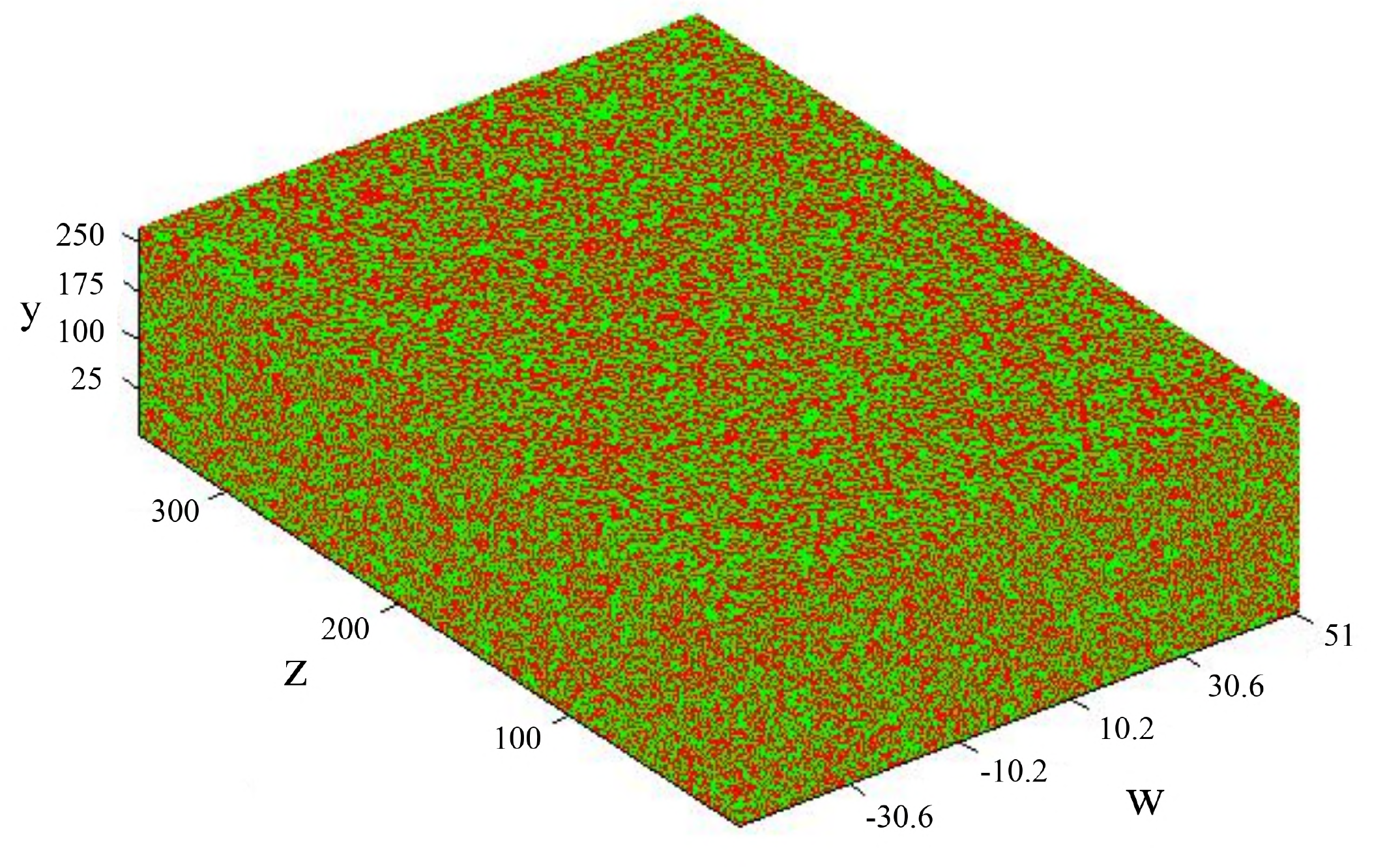}
\end{center}
\caption{Basins of attraction for attractors $A_1$ (red dots) and $A_2$ (green dots) on the Poincar\'e map ($x=0$) for $r=453$. The other parameters are as in Fig. \ref{fig bifdiag}.}
\label{fig basins}
\end{figure}

\begin{figure}[!htp]
\begin{center}
\includegraphics[width=7.0cm]{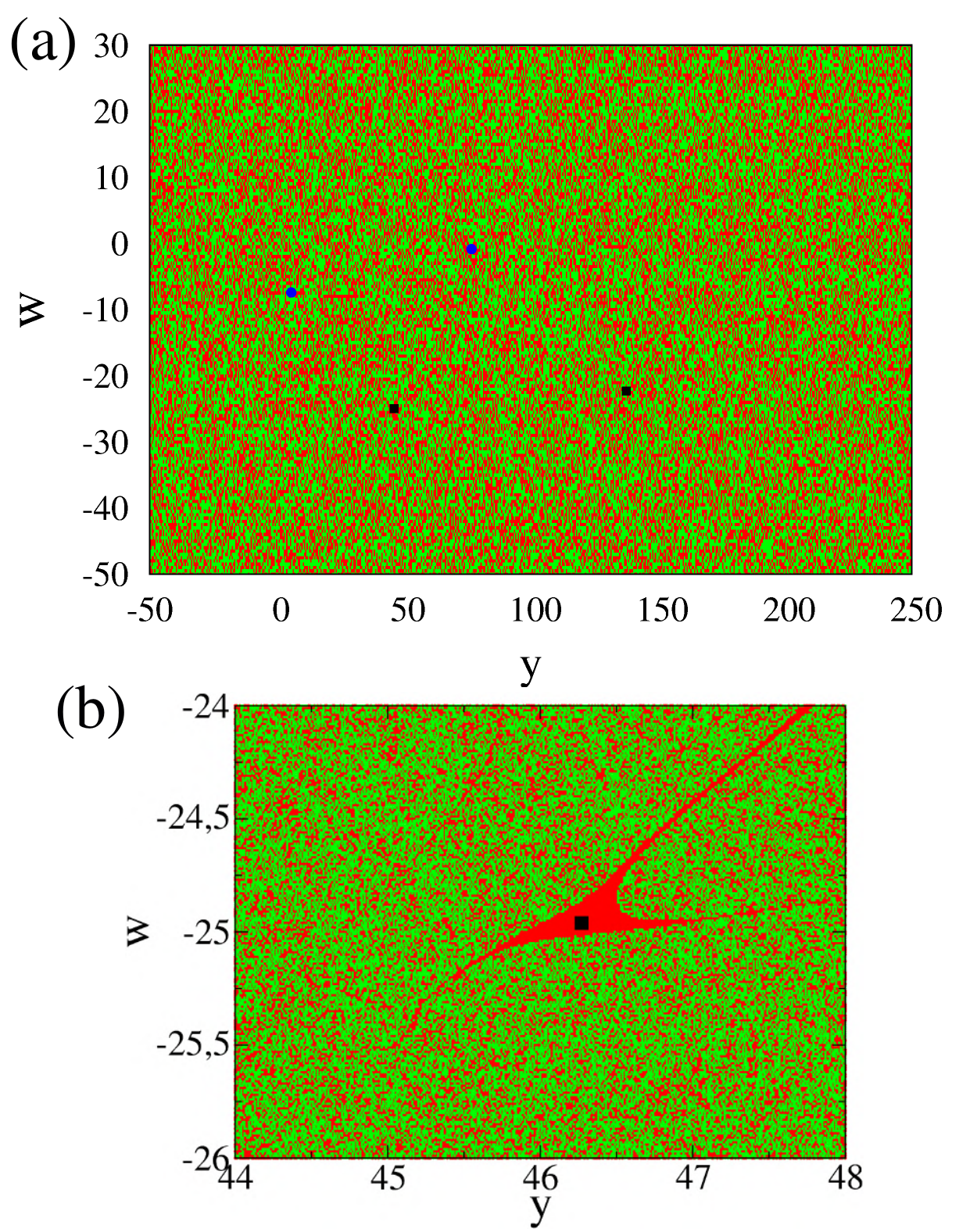}
\end{center}
\caption{(a) A two-dimensional slice $(w \times y)$ of the basins of attraction shown in Fig. \ref{fig basins}. Black squares represent the period-2 attractor $A_1$ and blue circles represent the period-2 attractor $A_2$. Red dots converge to $A_1$ and green dots converge to $A_2$. (b) An enlargement of a region in (a), near one of the points of $A_1$.}
\label{fig basins_zoom}
\end{figure}

Inside the periodic window, $A_1$ and $A_2$ attract different sets of initial conditions, defined as basins of attraction. The boundary between these basins is highly complex, as illustrated by Fig. \ref{fig basins} for $r=453$, where red dots represent initial conditions whose trajectories eventually converge to $A_1$ and green dots represent initial conditions that converge to $A_2$. The complexity is scale invariant, as successive amplifications of a region in the phase space do not simplify the picture, as confirmed by Fig. \ref{fig basins_zoom}(a), where a two-dimensional slice of the phase space is shown nearby the period-2 attractors $A_1$ (black squares) and $A_2$ (blue circles). These points represent the Poincar\'e points of the symmetric attracting trajectories shown in Fig. \ref{fig symmetry}, but in a different projection $(w \times y)$. Figure \ref{fig basins_zoom}(b) shows an enlargement of a region around one of the Poincar\'e points of $A_1$, where it is clear that the basin only becomes smooth in the close vicinity of the attractor, with an apparently fractal structure otherwise.

\begin{figure}[!htp]
\begin{center}
\includegraphics[width=7.5cm]{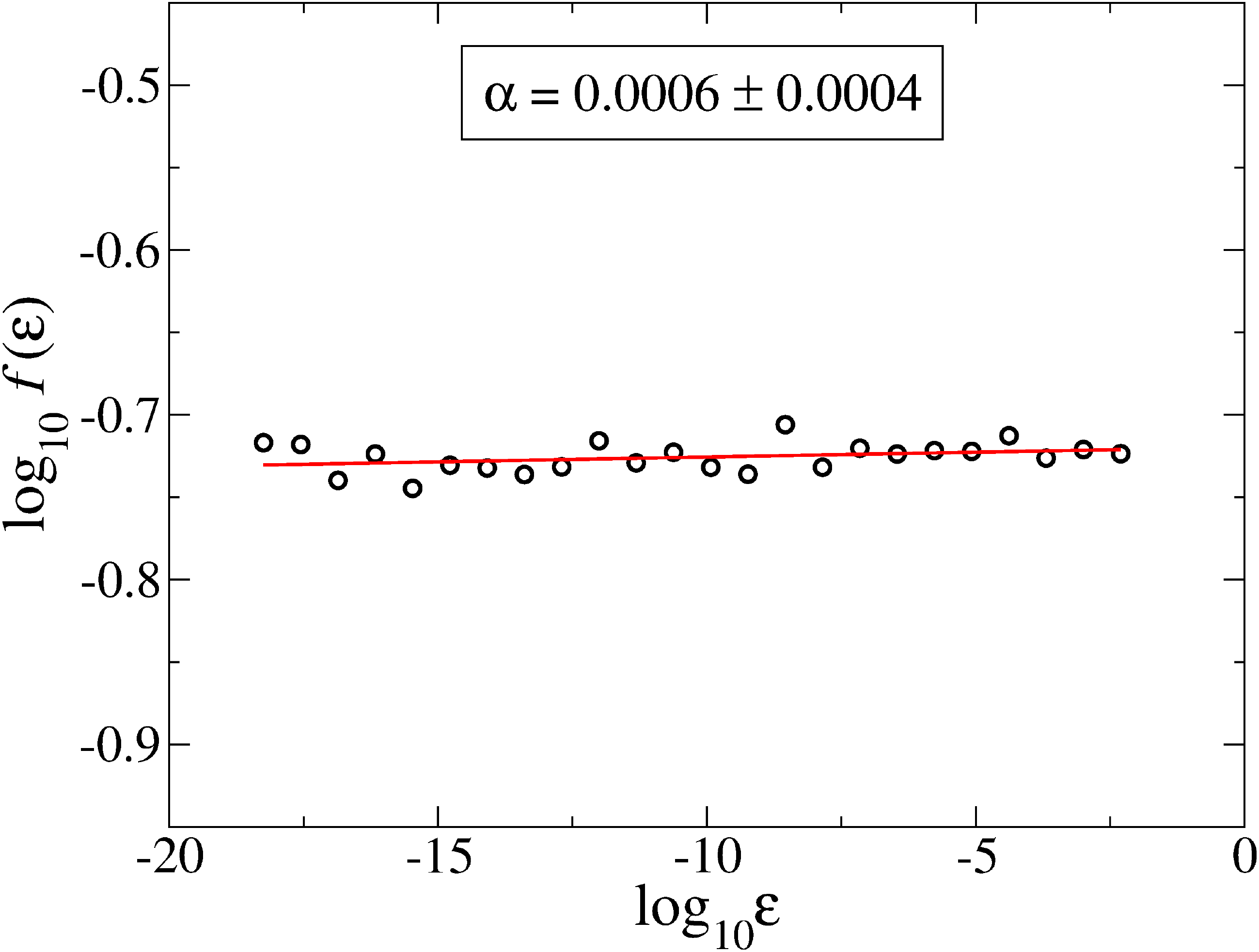}
\end{center}
\caption{Loglog plot of the fraction of uncertain initial conditions $f$ as a function of the uncertainty parameter $\varepsilon$ for $r=453$. The slope of the fitted line is $\alpha = 6\times 10^{-4} \pm 4\times 10^{-4}$.} 
\label{fig f}
\end{figure}

The fractal dimension of the basin boundary can be estimated with the aid of the uncertainty exponent \cite{grebogi1983}. Start with a line connecting two points in the phase space, arbitrarily chosen. Then, randomly choose $N_s=10000$ initial conditions on this line and determine to which basin of attraction each of them belongs. Next, displace each initial condition by adding a small perturbation $\varepsilon$. A point is considered uncertain if its perturbation converges to a different attractor. Compute the number of uncertain points $N(\varepsilon)$ for different values of $\varepsilon$ and obtain the fraction of uncertain points as

\begin{equation}
f(\varepsilon)=\frac{N(\varepsilon)}{N_s}.
\end{equation}
In fractal basin boundaries, the fraction scales as $f \sim \varepsilon^\alpha$, so $\alpha$ is the slope of the linear relation between $\log{(f)}$ and $\log{(\varepsilon)}$. The graph of $\log{(f)} \times \log{(\varepsilon)}$ is shown in Fig. \ref{fig f}, from which the slope of the linear regression is $\alpha = 6\times10^{-4}\pm4\times10^{-4}$. The dimension of the set of intersecting points of the basin boundary with the 1D line is $d_s = 1 - \alpha$. In the full 3D Poincar\'e map, the dimension of the basin boundary is $D_s = 3-\alpha = 2.9994$, a value extremely close to the dimension of the phase space.  

\begin{figure}[!htp]
\begin{center}
\includegraphics[width=10.0cm]{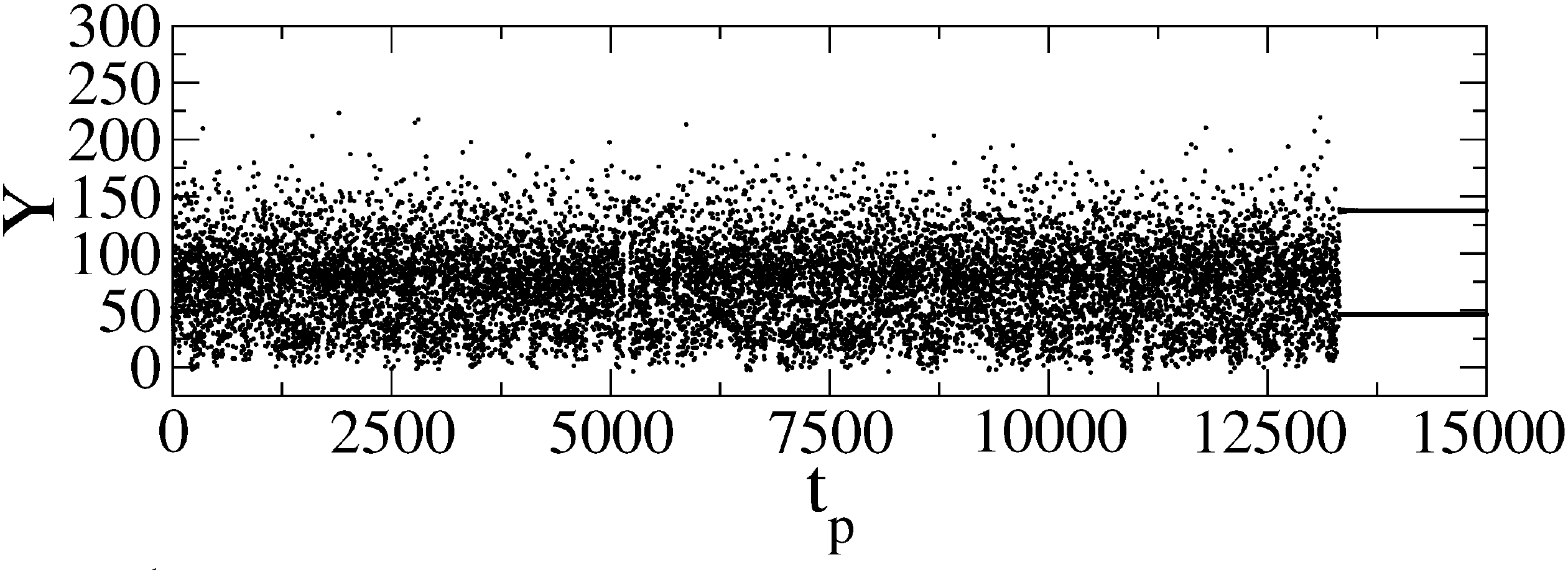}
\end{center}
\caption{Long chaotic transient before convergence to a period-2 attractor in the Poincar\'e map for $r=453$.} 
\label{fig transient}
\end{figure}

The intricacy of the basin boundary and its high fractal dimension result in long chaotic transients before the solutions settle to an attractor. Figure \ref{fig transient} shows a solution with a chaotic behaviour up to $t_p \approx 13600$ iterations of the Poincar\'e map for $r=453$, before the system converges to a period-2 attractor.

\begin{figure}[!htp]
\begin{center}
\includegraphics[width=7.5cm]{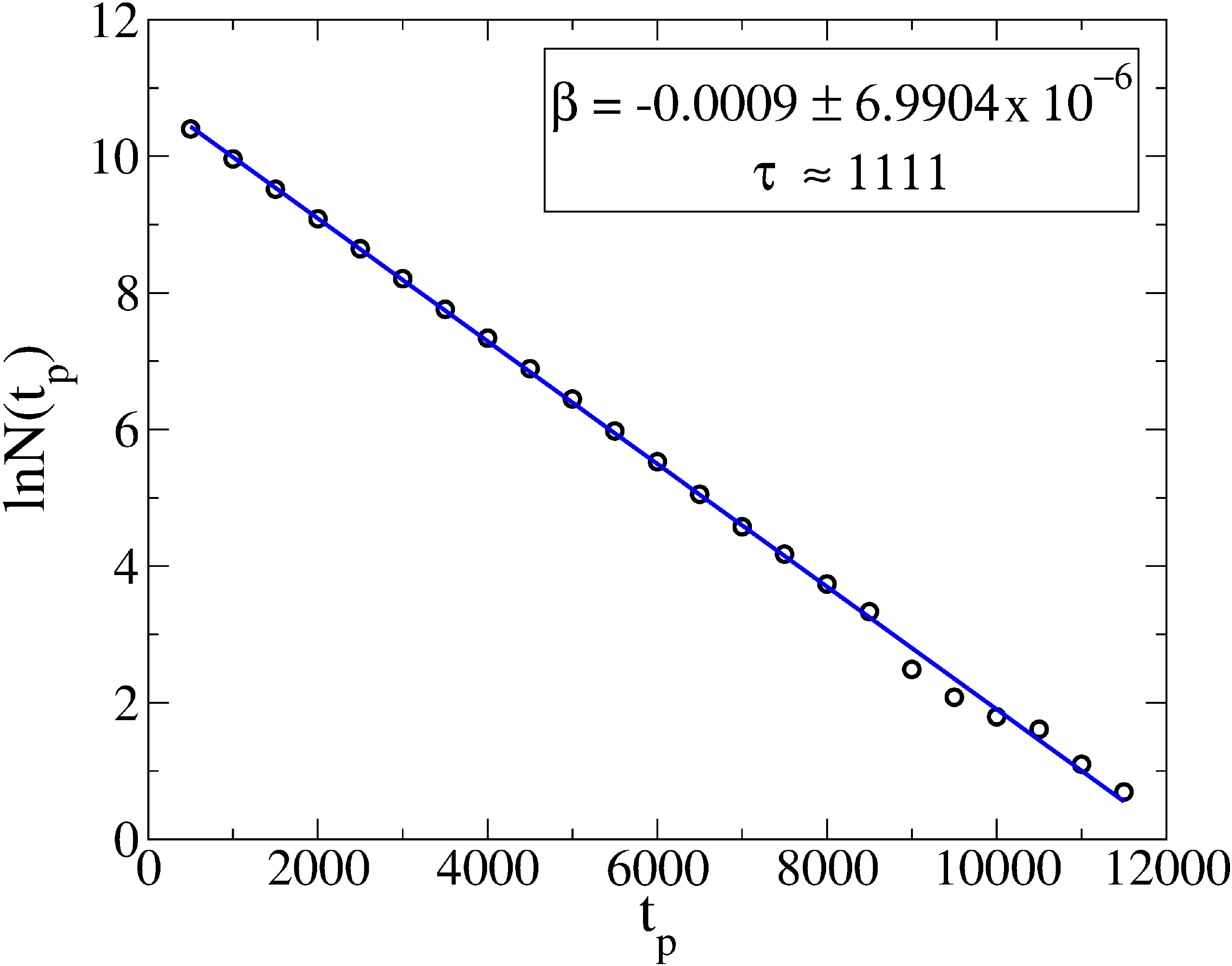}
\end{center}
\caption{Log-linear plot of $N(t)$, the number of orbits that have not converged to an attractor after $t_p$ Poincar\'e map iterations, as a function of $t_p$ for $r=453$. 
The other parameters are as in Fig. \ref{fig bifdiag}. The slope of the fitted line is $\beta = -9\times 10^{-4} \pm 6.5 \times 10^{-6}$, providing an average transient time of $\tau \approx 1111$ Poincar\'e map iterations.}
\label{fig Nt}
\end{figure}

As mentioned before, these chaotic transients are due to the presence of a chaotic saddle in the phase space. Inside the periodic window, this chaotic saddle $\Lambda_s$ is located at the boundary between the basins of attractors $A_1$ and $A_2$, which coincides with the stable manifold of the chaotic saddle \cite{pentek1995,tel2015}. In order to find $\Lambda_s$, we first determine the average transient time $\tau$ of initial conditions in the phase space. We define the lifetime of a trajectory as the time it takes to go from its initial condition until the close vicinity of an attractor. For $r=453$, we define a $500 \times 100$ grid of initial conditions in the $(w\times y)$ plane, with the other state variables fixed. Let $N_0$ be the number of initial conditions in the grid and let $N(t)$ be the number of trajectories from those initial conditions that have not converged to any attractor after $t$ iterations of the Poincar\'e map. These trajectories must be near the chaotic saddle $\Lambda_s$ and, due to its chaotic nature, the probability that the trajectory has not yet escaped from the vicinity of the chaotic saddle on time $t$ decays exponentially with time ~\cite{Kantz1985}

\begin{equation}
P(t) \sim exp(-\kappa t),
\end{equation}
for some $t>t_0$, where $\kappa$ is the decay rate. Following ~\citet{hsu1988, lai1995, sweet2000}, we write $P(t) = N(t)/N_0$ and $\kappa = 1/\tau$. 

\begin{equation}
N(t) \sim N_0\exp(-t/\tau).
\label{eq N}
\end{equation}
Equation (\ref{eq N}) defines a linear dependence between $\log{N(t)}$ and $t$, with $-1/\tau$ as the slope. Thus, the average transient time $\tau$ can be computed by considering the inverse of the slope of the graph of $\log(N(t))\times t$, as shown in Fig. \ref{fig Nt}. The estimated value from linear regression is $\tau \approx 1111$.

\begin{figure}[!htp]
\begin{center}
\includegraphics[width=7.5cm]{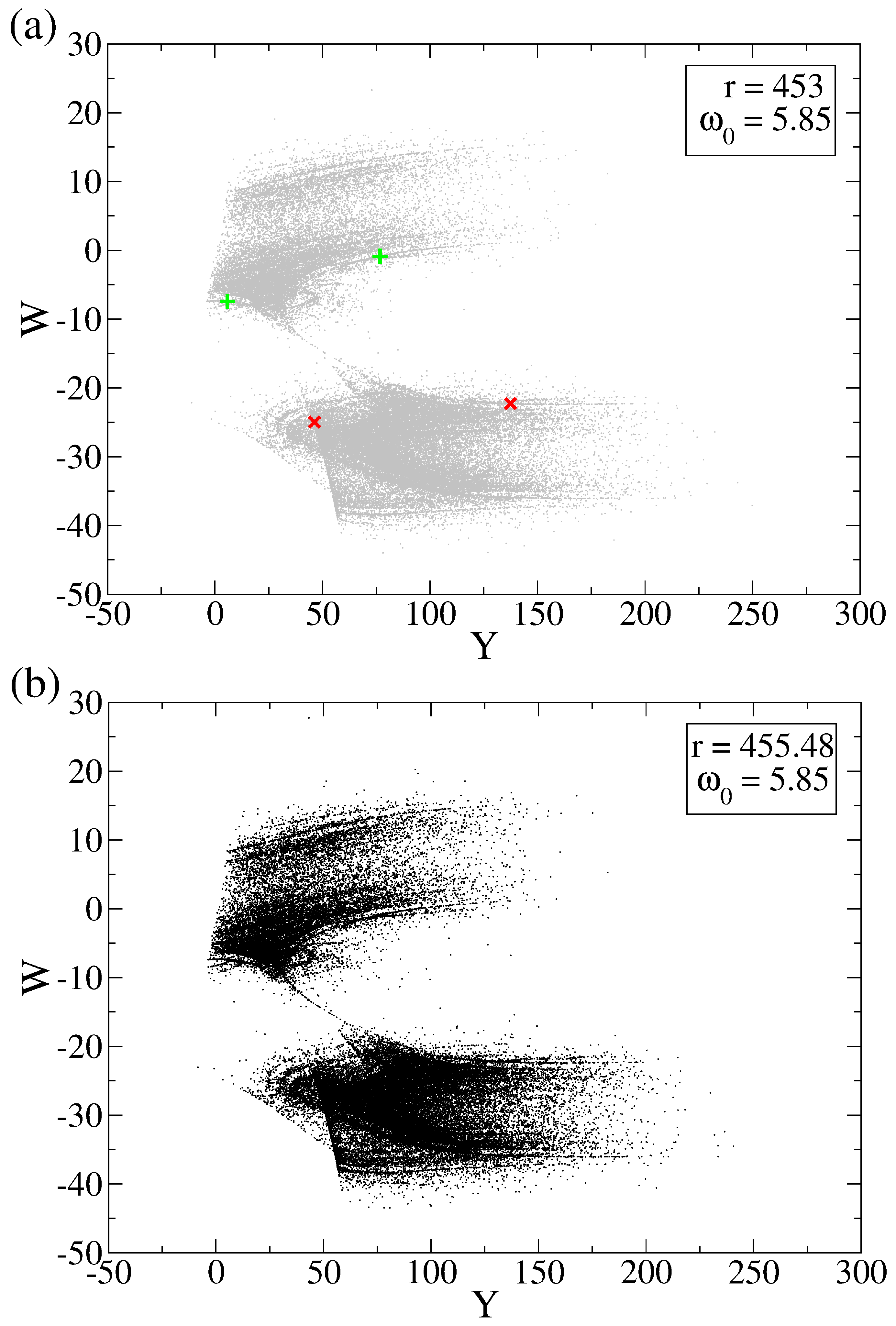}
\end{center}
\caption{(a) Poincar\'e points of the $(w \times y)$ components of the chaotic saddle inside the periodic window at $r=453$. The other parameters are as in Fig. \ref{fig bifdiag}. Red crosses represent attractor $A_1$ and green crosses represent $A_2$. (b) The chaotic attractor at $r=455.48$}
\label{fig cs}
\end{figure}

We now describe how the sprinkler method is used to find $\Lambda_s$. First, we find on a grid the set of initial conditions with trajectories that are still chaotic after a long time $t_c$. The value of $t_c$ must be large compared to the average lifetime $\tau$, so we choose $t_c=3000$. These initial conditions will first approach $\Lambda_s$ through its stable manifold, stay in its vicinity for some time before they depart along the unstable manifold toward an attractor. Thus, if we iterate all selected initial conditions until $t_m=t_c/2$, their trajectories must be very close to $\Lambda_s$. The set of all those points approximate the chaotic saddle. Figure \ref{fig cs}(a) depicts the $(y\times w$) components of the chaotic saddle in the beginning of the periodic window, at $r=453$, to the left of SNB in Fig. \ref{fig bifdiag}(a), when the attractors are periodic. The period-2 attractors are plotted as red ($A_1$) and green ($A_2$) crosses. A comparison with the chaotic attractor to the right of SNB, shown in Figure \ref{fig bifdiag}(b) for $r=455.48$, reveals that the chaotic saddle is formed by a continuation of many of the recurrent points found in the pre-window chaotic attractor, as discussed in \citet{robert2000}. 

\begin{figure}[!htp]
\begin{center}
\includegraphics[width=7.5cm]{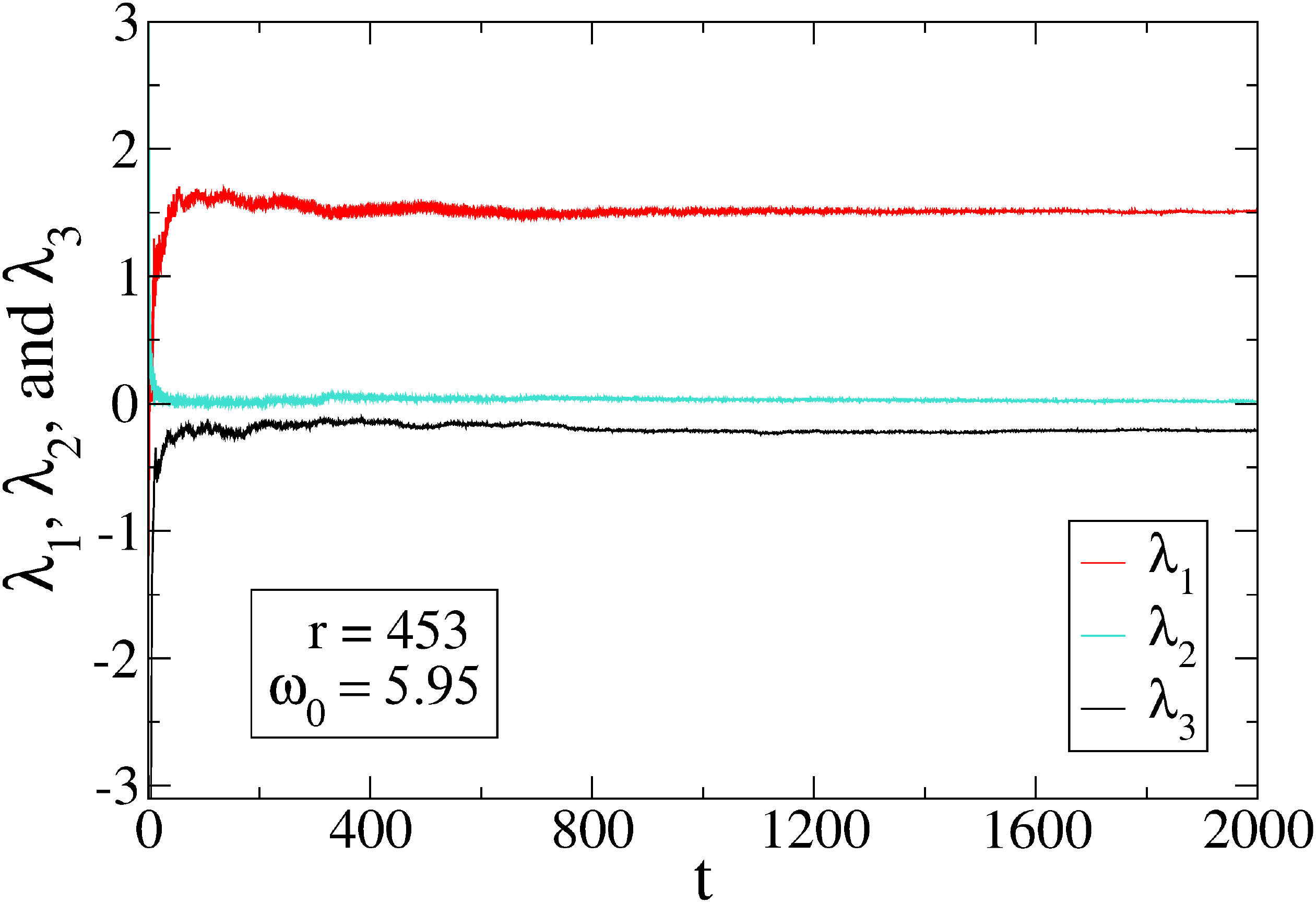}
\end{center}
\caption{Convergence of the first three Lyapunov exponents of a long chaotic transient for $r=453$.}
\label{fig lyapcs}
\end{figure}

The Lyapunov exponents of the chaotic saddle $\Lambda_s$ can be approximately computed
as the Lyapunov exponents of a long chaotic transient. Figure \ref{fig lyapcs} shows the convergence of the first three Lyapunov exponents for a long chaotic transient at $r=453$. We can compare the value $\lambda_1\approx 1.5$ of the chaotic saddle with the maximum Lyapunov exponent of the chaotic attractor just to the right of SNB in Fig. \ref{fig bifdiag}(b). For $r=455.58$, $\lambda_1\approx 1.51$ for the chaotic attractor, a value very close to the maximum Lyapunov exponent of $\Lambda_s$, confirming the relation of continuation between both sets. 

The maximum Lyapunov exponent $\lambda_1$ of the chaotic saddle can also be estimated from the average lifetime $\tau$ and from $d_s$, the fractal dimension of the set of intersecting points of a one-dimensional line with the stable manifold of the chaotic saddle. For two-dimensional maps, this relation is given by \cite{hsu1988}

\begin{equation}
\tau = 1/[(1-d_s)\lambda_1].
\label{eq tau}
\end{equation}
Note that \citet{Kantz1985} had previously derived a more general relation for the decay rate as $\kappa = (1-D_1) \lambda_1$, where $D_1$ is the partial information dimension of the chaotic saddle. For a review of this topic, see the chapter 8 of ~\citet{Lai2011}.  

For certain higher-dimensional phase spaces, it has been argued that the same relation holds \cite{lai1995}. Using the previously computed values of $\tau$ and $d_s$ in Eq. (\ref{eq tau}) for $r=453$, we obtain 

\begin{equation}
\lambda_1=1/[(1-(1-\alpha))1111].
\label{eq l}
\end{equation}
The error bar in $\alpha=6\times 10^{-4} \pm 4\times 10^{-4}$ is too large for a precise estimation of $\lambda_1$ with Eq. (\ref{eq l}). Nonetheless, we observe that using the mean value $\alpha=6\times 10^{-4}$ yields 

\begin{equation}
\lambda_1=1/[(1-0.9994)1111] \approx 1.5.
\label{eq l2}
\end{equation}
The value agrees quite well with the one computed directly from the chaotic transient in Fig. \ref{fig lyapcs}, confirming that the chaotic transients are due to the chaotic saddle localized on the fractal basin boundary. 

\section{Discussion and Conclusions}
\label{sec conclusion}

Transient chaos is still an overlooked topic in most textbooks and theoretical works on dynamical systems, where the tendency is to focus exclusively on the asymptotic dynamics, after trajectories have converged to an attractor. In practice, however, it is usually not possible to proof that an observed behaviour is asymptotic or not, as experimental results can only confirm a certain behaviour up to a finite time scale, as noted by \citet{tel2015}. This has strong implications in areas such as transition to turbulence in pipe flows, where long chaotic transients have been observed and it is difficult to find a critical Reynolds number where the system switches from laminar to persistent turbulence and chaotic saddles play a crucial role \cite{eckhardt2007}. In this case, the lifetime of the transient chaos follows a supertransient law, i.e., it grows exponentially as a function of the control parameter \cite{tel2008}, a phenomenon also observed in numerical simulations of Keplerian shear flows in the context of accretion disks \cite{rempel2010,danilo2018}. For other applications of transient chaos, including intermittency and scattering in leaking systems, see \citet{tel2008}, \citet{altmann2013} and \citet{tel2015}.  

Our results illustrate the importance of transient chaos in dynamical systems in general and in this magnetoconvection model in particular. The existence of very long chaotic transients as seen in Fig. \ref{fig transient} can mask the true attractors of the system if the time scales considered are smaller than the average transient time. In the case of the Lorenz system, we stress that the presence of a magnetic field causes an increase in the transient lifetime in the range of parameters studied in this paper. It is also worth pointing that the original Lorenz model displayed multistability, with up to three coexisting basins of attraction reported by \citet{yorke1979}, but the basin boundaries are not fractal in those cases \cite{sprott2017}. We are unaware of the presence of basin boundaries with a fractal dimension close to the dimension of the phase space (as seen in Figs. \ref{fig basins} and \ref{fig basins_zoom}) in the original Lorenz system. The high fractal dimension is directly related to the long transients reported above, as indicated by Eq. (\ref{eq tau}).

Regarding the non-dimensional control parameters, they were chosen according to the values commonly employed in the original Lorenz model to focus on the impact of the addition of a magnetic field. The new parameters are $\omega_0=5.95$, responsible for the intensity of the background magnetic field, and the magnetic Prandtl number $\sigma_m=0.1$, whose values were chosen following \citet{macek2014}. Although it is a low $\sigma_m$, this value is orders of magnitude higher than what is found in the Earth's liquid outer core ($\sigma_m\sim 10^{-6}$) or in stellar interior ($10^{-4} \leq \sigma_m \leq 10^{-6}$) \cite{mondal2018}. In fact, considering the relevance of the present work for the study of magnetoconvection, we don't claim that Macek's reduced Lorenz model provides realistic simulations of magnetized convection in stellar or Earth's interior, where this phenomenon is responsible for the dynamo that maintains the magnetic fields \cite{weiss2014}. However, local and global bifurcations, chaotic saddles and transients similar to the ones displayed by this model have also been observed in more realistic direct numerical simulations of three-dimensional Rayleigh-B\'enard convection \cite{roman2015}. Thus, the analysis of the reduced model can shed light to the understanding of the complexity present in magnetoconvection and guide future nonlinear analyses of more realistic models. 

\section*{Acknowledgments}
This work had the financial support of Brazilian funding agencies CAPES (88887.309065/2018-00), CNPq (304449/2017-2) and FAPESP (2013/26258-4).

\bibliographystyle{unsrt}


\end{document}